\documentclass[12pt]{article}
\input epsf.sty
\newcommand{\nl}{\hspace{-.65cm}}
\newcommand{\be}{\begin{equation}}
\newcommand{\ee}{\end{equation}}
\newcommand{\ben}{\begin{eqnarray}\displaystyle}
\newcommand{\een}{\end{eqnarray}}

\def\sqr#1#2{{\vcenter{\vbox{\hrule height.#2pt
         \hbox{\vrule width.#2pt height#1pt \kern#1pt
            \vrule width.#2pt}
         \hrule height.#2pt}}}}

\begin{document}

\begin{center}
\large{\bf String Theory Versus Black Hole Complementarity }

\vspace{10mm}

\normalsize{Amit Giveon \\\vspace{2mm}{\em Racah Institute of Physics, The Hebrew University, Jerusalem, 91904, Israel} \\\vspace{2mm}{\em and} \\\vspace{2mm}  Nissan Itzhaki \\\vspace{2mm}{\em Physics Department, Tel-Aviv University,
Ramat-Aviv, 69978, Israel}}


\end{center}

\vspace{10mm}

\begin{abstract}

It is argued that string theory on the Euclidean version of the Schwarzschild black hole -- the cigar geometry -- admits a zero mode that is localized at the tip of the cigar. The presence of this mode implies that in string theory, unlike in general relativity, the tip of the cigar is a special region. This is in tension with the Euclidean version of the black hole complementarity principle. We provide some qualitative arguments that link between this zero mode and the origin of the black hole entropy and firewall at the horizon.

\end{abstract}

\newpage

\baselineskip=18pt


\newpage
\bigskip

A key question to the understanding of quantum black holes is whether the black hole horizon of a large BH acts just like a Rindler horizon. For example, the Unruh radiation that a Rindler observer measures does not curve space-time.  Does this mean that  the curvature at the horizon of a large black hole remains small at the quantum level?
The Black Hole Complementarity (BHC) principle \cite{Susskind:1993if} postulates that the answer is in the affirmative. Arguments against this postulate were put forward sometime ago \cite{Itzhaki:1996jt} in the context of 't Hooft S-matrix for the black hole \cite{'tHooft:1996tq}, and  more recently  using  quantum information theory in \cite{Almheiri:2012rt}. It is natural to wonder what   string theory has to say   about this  question.
In this short note we argue, in the context of Euclidean black holes,  that already at the perturbative level string theory provides concrete evidence against BHC.

Consider string theory on the Euclidean version of the black hole -- the cigar background. For concreteness, we focus on the cigar background  obtained by Wick rotating the Schwarzschild solution in four dimensions. Our conclusions, however, are rather general. The solution reads \footnote{We ignore the compact manifold which plays no role here.}
\be
ds^2=\left(1-\frac{2GM}{r}\right)dx_4^2+\frac{dr^2}{1-\frac{2GM}{r}}+r^2d\Omega_2^2,
\ee
with $x_4\sim x_4 +8\pi G M$.

The horizon of the Schwarzschild solution is mapped to the tip of the cigar (at $r=2GM$). In General Relativity (GR) nothing special happens at the tip. The periodicity of $x_4$ ensures that there is no singularity there. For a large black hole the tip looks just like a flat space ($R^4$ in our case).  This is in accord with the Euclidean version of the BHC principle.

Can string theory change this conclusion? Naively, the answer is no. It is true that on top of the gravitons of GR string theory has other massless modes. But non of these modes is sensitive to the tip of the cigar. Moreover, integrating out the massive modes yields the $\alpha'$ corrections that, much like the stringy loops corrections, are small for smooth backgrounds. Hence, it seems hard to imagine how string theory can modify the GR conclusion.

This kind of reasoning is local by nature, and indeed there is nothing special {\em locally} about the  the tip of the cigar.
Globally, however, the tip of the cigar is special. It is the only point invariant under translation in $x^4$. GR is a local theory and hence is not sensitive to such global considerations. String theory is non-local and could potentially  be sensitive to this global aspect of the cigar geometry.

We now argue that indeed  string theory is sensitive to the global structure of the cigar  in a rather interesting way. The mode spectrum of the theory is fixed at infinity. At infinity the radius of the sphere becomes infinite and the 2d geometry described by $r$ and $x^4$ is of a cylinder with a large but finite radius ($R_{\infty}=4GM$). The GSO projection in such a case keeps tachyon modes with odd windings on the $x^4$ circle. Far from the tip these modes are extremely massive and contribute very little to the stringy corrections to the cigar background (even less than the usual $\alpha'$ corrections). However, as we approach the tip they become lighter and  lighter. At the tip itself they are formally tachyonic \cite{Kutasov:2005rr} and could become the most relevant degrees of freedom.

We would like to show now that indeed this is the case. To emphasis the fact that this conclusion is not correlated in any way with the small curvature of the geometry  but is only due to the GSO projection fixed at infinity, we take  the near tip approximation in which the background is flat,
\be\label{or}
ds^2=d\rho^2+\rho^2 d\phi^2 +dx_1^2 +dx_2^2.
\ee
The origin of $R^2$ is the tip of the cigar and $\phi=x^4/4GM$, so we have the usual periodicity $\phi\sim\phi+2\pi$.
To keep the horizon size finite, we compactify  the transverse directions as well, $x_1\sim x_1+L,~~x_2\sim x_2+L$, so  $A=L^2$ plays the role of the horizon area.

This is a  standard background in string theory. What is less standard is the list of modes that propagate on this flat background.  The modes are fixed not by the usual GSO projection rather by the GSO projection inherited from the cigar.  In particular it keeps tachyon modes with odd winding number.  Thus this GSO projection breaks translation invariance and picks $\rho=0$ as a preferred point. If all the modes that are sensitive to the origin of $R^2$ were heavy than this effect was interesting but
likely to induce negligible corrections to the physics of massless modes at the tip.
We now argue that in fact these new modes contain zero modes that are localized at the tip. The presence of such  zero modes implies that in string theory the tip of the cigar is special even for massless modes. This is in clear tension with the Euclidean version of the BHC principle,
which states that the tip of the cigar of a large black hole is equivalent to $R^2$.

The effective action that describes the winding modes (that now wind around $\phi$) is~\cite{Kutasov:2005rr}
\be\label{ac}
S=\frac12 \int_{0}^{\infty}d\rho \rho\left( (\partial_{\rho} T)^2 +m^2(\rho) T^2 \right),~~~~~~~~m^2(\rho)=(-1+w^2\rho^2/4),
\ee
where $w=2n+1$ is the winding number and we work with $\alpha'=2$.
We see that near the origin $m^2$ becomes negative \cite{Kutasov:2005rr}. Because of  the kinetic term  this by itself does not mean that these modes condense. In fact, one can show that for $w>1$ all configurations have positive action.

For $w=\pm 1$ something rather interesting happens. There is a non-trivial solution to the equation of motion with $S=0$. Namely, there is a zero mode,\footnote{One can show that despite of the $-T^2$ term in the action there are no configurations with $S<0$.}
\be
T(\rho)=\exp(- \rho^2/4).
\ee
From a CFT point of view, this zero mode was found in the context of NS5-branes by Kutasov and Sahakyan \cite{Kutasov:2000jp}. They found it to exist for arbitrarily small curvature and string coupling constant. In such a case, the tip of the NS5-branes looks locally just like
(\ref{or}), and the crucial point is again the non-trivial GSO projection that is determined globally.

The presence of this zero mode that is confined to the tip leads us to conclude that, unlike in GR, in string theory the tip of the cigar  is special. Quantum mechanically, localized  zero modes are bound to fluctuate and affect  the dynamics  at the  tip of the cigar.

This is a rather intriguing zero mode. Usually, a bosonic zero mode is associated with a collective coordinate that parameterizes the Euclidean solution. This does not appear to be the case here. The cigar is a solution of GR while this zero mode describes a degree of freedom that does not exist in GR. Moreover, the existence of this zero mode looks like a technical coincidence due to relations between parameters in string theory -- it depends on the $1/4$ in $m^2(\rho)$, which in turn depends on the dimensionless parameter that relates the mass of the tachyon with
the string tension.

There are  examples both in field theory \cite{Seiberg:1994rs} and string theory \cite{Strominger:1995cz}
in which a field that on  generic points in the manifold  is heavy and  contributes very little to the low-energy dynamics becomes light at a special point where it takes over the dynamics. This special point is singular without this field and by becoming light  the field resolves the singularity.
At first it seems like our example is different since there is no singularity at the tip of the cigar in GR.
However, in string theory the GSO projection, fixed at infinity, is such that the modes are classified by the winding number. At the tip the size of these winding goes to zero and so this classification leads to a singularity. This singularity is resolved by the winding zero modes that smear it.
So in a sense string theory introduces a singularity and its resolution at the same time.

We now argue that these zero modes  might play an important  role in the microscopic understanding of the black hole entropy.

The classical action associated with $R^2 \times T^2$ vanishes.
We wish to estimate the contribution of the localized modes to the one-loop  action. As usual, to do so we have to consider the quantum fluctuations
eigenvalue equation $
M\Phi= E \Phi$,
which gives
\be
S_1=\log(\det M)= \mbox{Tr} \log(M).
\ee
In our case,
\be
M=\frac{1}{\rho}\partial_{\rho}+\partial_{\rho}^2+\partial_{1}^2+\partial_2^2
-(-1+\rho^2/4),
\ee
and the relevant modes are
\be
\Psi(n_1,n_2)\sim \exp(- \rho^2/4) \exp\left(\frac{i}{L}(n_1 x^1+n_2 x^2)\right),
\ee
with eigenvalues  $ (n_1^2+n_2^2)/L^2$.
Thus we get
\be
S_1\sim A \Lambda^2,
\ee
where $\Lambda $ is the cutoff on the transverse momentum. It is trivial yet interesting in relation to black hole entropy that $S_1$ is linear in $A$. The question is what is $\Lambda$? Normally, in string theory we do not have to introduce a UV cutoff and the string scale appears naturally in the calculation. This is clearly the case for the standard modes on (\ref{or}), which propagate all over space and give $V_4/l_s^4$.

While it is reasonable to suspect that modular invariance will impose that also for the new modes $\Lambda$ is the string scale, and so $S_1\sim A/l_s^2$, it is less clear that this must be the case since here we are talking about excitations of zero modes that are not related to collective coordinates. Below we present a naive  argument which suggests  that for $S_1$ the relevant cutoff scale is the Planck scale and not the string scale.

One way to estimate $\Lambda$ is to note that the form of the  corrections to the free field action (\ref{ac})  scales like
\be
g_s^2 T^4.
\ee
This implies that our free field theory approximation breaks down when $T^2\sim 1/g_s^2$. Since each normalizable mode scales like $1/L$ and the modes are not correlated, we get $T^2 \sim \frac{N}{L^2}$, where $N$ is the total number of modes $L^2 \Lambda^2$. This means that $\Lambda \sim \frac{1}{g_s}$ and that
\be
S_1 =c \frac{A}{G},
\ee
where $c$ is a numerical coefficient of order $1$ we cannot fix. This {\em naive} estimate suggests that these modes could play a key role in the microscopic understanding of the black hole entropy. Needless to say it would be nice to explore this further.

We would like to conclude by making contact with the result of  \cite{Almheiri:2012rt}.  In \cite{Almheiri:2012rt} it was shown that if Hawking radiation is in a pure state then at least for old black holes a Hawking particle and its partner cannot be in a pure state.
Namely, their entanglement entropy cannot vanish,
\be\label{hp}
S_{Haw-Par}>0.
\ee
This conclusion follows from subadditivity of entanglement entropy and as such is hard to avoid.

In Euclidean space, the statement that a pair is created in a pure state is the fact that this process is described by a bubble diagram that closes without external interactions. This is exactly what happens in ordinary $R^2$ which is the reason why an Unruh particle and its partner form a pure state (and there is no firewall at the Rindler horizon).

In our case, we have $R^2$ but with large fluctuations of the tachyonic field at the origin. In such a situation most bubbles that are away from the origin will close with no real  interaction with the localized tachyonic modes. These bubbles, however, are not the analog of a Hawking particle (that makes it all the way to infinity) and its partner (that falls towards the singularity).
The analog of such a pair is a bubble that almost crosses the origin. Such a bubble interacts with the localized modes, and it is thus consistent with the conclusion that (\ref{hp}) has to be satisfied. This suggests that the firewall is made out of fundamental strings that are stretched all over the horizon along the null Killing direction (which is the analog of  $\phi$ in (\ref{or}) at $\rho=0$). The cigar is the analog of an infinitely old black hole. Hence we cannot address the question raised in \cite{Susskind:2012ey} whether  it is  the  scrambling time or the Page time that is relevant for the formation of the firewall.

The fact that fundamental strings play a key role in the dynamics of black hole horizons has a long history going back to the work of Susskind \cite{Susskind:1993ki}.  Moreover, the fact that strings tend to stretch along the null Killing direction at the horizon was also pointed out long ago \cite{Susskind:1993aa}. We hope that  input from the  Euclidean setup will provide some useful clues for the more challenging Lorentzian system.

\vspace{10mm}

\nl{\bf Acknowledgments}\\
We thank Ofer Aharony, Ramy Brustein, Shmuel Elitzur and David Kutasov for discussions.
The work of AG is supported in part
by the BSF -- American-Israel Bi-National Science Foundation,
and by a center of excellence supported by the Israel Science Foundation
(grant number 1665/10).
The work of NI is supported in part by the Israel Science Foundation (grant number 1362/08) and by the European Research Council (grant number 203247).

\end{document}